\documentstyle [12pt] {article}
\begin{document}
\setcounter{page}{1}
\def\theequation{\arabic{section}.\arabic{equation}}
\def\theequation{\thesection.\arabic{equation}}
\setcounter{section}{0}

\title{On the $D^*D\,\pi$ and $\bar{B}^* \bar{B} \pi$ coupling constants\\
in the effective quark model with chiral symmetry}

\author{A. N. Ivanov\thanks{E--mail: ivanov@kph.tuwien.ac.at, Tel.:
+43--1--58801--5598, Fax: +43--1--5864203}~{\footnotesize$^{^{\dag}}$},
N.I. Troitskaya\thanks{Permanent Address:
State Technical University, Department of Theoretical
Physics, 195251 St. Petersburg, Russian Federation}}

\date{}

\maketitle

\begin{center}
{\it Institut f\"ur Kernphysik, Technische Universit\"at Wien, \\
Wiedner Hauptstr. 8-10, A-1040 Vienna, Austria}
\end{center}

\vskip1.0truecm
\begin{center}
\begin{abstract}
The coupling constants $g_{D^{\ast}D\pi}$ and
$g_{\bar{B}^{\ast}\bar{B}\pi}$ of the $D^{\ast}D \pi$ and
$\bar{B}^{\ast}\bar{B}\pi$ strong interactions are computed in the
effective quark model with chiral symmetry incorporating Heavy quark
effective theory (HQET) and Chiral perturbation theory at the quark level
(CHPT)$_q$ with linear realization of chiral $U(3)\times U(3)$ symmetry. We
predict  $g_{D^{\ast}D\pi} = 5.2$ and $g_{\bar{B}^{\ast}\bar{B}\pi} =
14.3$.
The $1/M_c$ corrections to the constant $g_{D^* D\pi}$ are calculated and
make up less than 1$\%$. The contribution of $1/M_b$ corrections to the
constant $g_{\bar{B}^*\bar{B}\pi}$ is much less.
\end{abstract}
\end{center}
\vspace{0.2in}

\begin{center}
PACS number(s): 13.30.Ce, 12.39.Ki, 14.20.Lq, 14.20.Mr
\end{center}

\newpage

\section{Introduction}
\setcounter{equation}{0}

The investigation of strong interactions $D^* D \pi$ and
$\bar{B}^*\bar{B}\pi$ runs parallel to the investigation of a dominance of
vector $D^*$ and $\bar{B}^*$ mesons in the semileptonic decays $D\to \pi
\ell \nu_{\ell}$ and $\bar{B}\to \pi \ell  \nu_{\ell}$ related to the
strong transitions $D\to \pi$ and $\bar{B}\to \pi$, respectively, the form
factors of which can be measured experimentally. In turn, the computation
of the coupling constant $g_{D^{\ast}D\pi}$ of the $D^* D \pi$ interaction
has a self--determined meaning, since the transitions $D^* \to D \pi$ do
exist on--mass-shell of interacting mesons. The probabilities of the decays
$B(D^{*+}\to D^0\pi^+) = (68.1\pm 1.3)\,\%$, $B(D^{*+}\to D^+\pi^0) =
(30.8\pm 0.8)\,\%$ and
$B(D^{*0}\to D^0\pi^0) = (63.6\pm 2.8)\,\%$ have been measured recently by
CLEO Collaboration [1].

The numerical values of $g_{D^{\ast}D\pi}$ and $g_{\bar{B}^*\bar{B}\pi}$
are distributed from $g_{D^{\ast}D\pi}=7\pm 2$, $g_{\bar{B}^*\bar{B}\pi} =
15\pm 4$ [2] and  $g_{D^{\ast}D\pi}=6.3\pm 1.9$, $g_{\bar{B}^*\bar{B}\pi} =
14\pm 4$ [3] up to $g_{D^{\ast}D\pi}=12.5\pm 1.0$, $g_{\bar{B}^*\bar{B}\pi}
= 29\pm 3$ [4]. For the more complete list of references to the numerical
values of the coupling constants $g_{D^{\ast}D\pi}$ and
$g_{\bar{B}^*\bar{B}\pi}$ we relegate readers to Table~1 of Ref.[4].
Most of these results have been obtained within QCD sum rules approach [5].

Recently [6] in the effective quark model with chiral $U(3)\times U(3)$
symmetry incorporating Heavy quark effective theory (HQET) [7--9] and
Chiral perturbation theory at the quark level (CHPT)$_q$ with linear
realization of chiral $U(3)\times U(3)$ symmetry [10] the coupling constant
$g_{D^{\ast}D\pi}$ has been computed at leading order in large $c$--quark
mass $M_c$ expansion with the following value $g_{D^{\ast}D\pi} =
g_{D^{\ast+}D^0 \pi^+} = 5.2\pm \Delta$. Here $\Delta$ is a theoretical
uncertainty of our effective quark model which on the estimate of [6] makes
up about 50$\%$. However, for all cases of the application of HQET,
supplemented by (CHPT)$_q$, to the description of strong interactions of
heavy--light mesons [11--16] the resultant agreements between theoretical
and experimental data turn out much better.

A possible substantial change of the $g_{D^{\ast}D\pi}$ coupling constant
value in HQET, supplemented by (CHPT)$_q$, one can expect at
next--to--leading order in large $M_c$ expansion. Indeed, as has been shown
in Ref.[15] such correcitions to the form factors of semileptonic
$D$--meson decays can make up more than 30$\%$.

This letter is to compute $g_{D^{\ast}D\pi}$ in HQET, supplemented by
(CHPT)$_q$, to next--to--leading order in large $M_c$ expansion and to
extend the obtained results on the computation of the coupling constant
$g_{\bar{B}^{\ast} \bar{B} \pi}$ defining the $\bar{B}^{\ast} \bar{B} \pi$
strong interaction.

However, we have found that the contribution of $1/M_c$ corrections to
$g_{D^{\ast}D\pi}$ makes up less than 1$\%$. In turn $1/M_b$ corrections,
where $M_b$ is the mass of the $b$--quark, to
$g_{\bar{B}^{\ast}\bar{B}\pi}$ is much less. The numerical value of
$g_{\bar{B}^{\ast} \bar{B} \pi}$ calculated at leading order in large $M_b$
expansion equals: $g_{\bar{B}^{\ast} \bar{B} \pi} = 14.3$. Thus, our
results turn out to be compared reasonably well with the predictions of
Refs.[2,3]. Therefore, in complete agreement with Refs.[2,3] we predict
much less constibutions of vector mesons $D^*$ and $\bar{B}^*$ to the form
factors of semileptonic decays  $D\to \pi \ell \nu_{\ell}$ and $\bar{B}\to
\pi \ell \nu_{\ell}$ than it can be expected due to Ref.[4].

\section{Calculation of coupling constants at leading order in heavy quark
mass expansion}
\setcounter{equation}{0}

The coupling constant $g_{D^{\ast}D\pi}=g_{D^{\ast +}D^0\pi^+}$ of the
$D^{\ast+}\to D^0 \pi^+$ decay is defined by the matrix element
\begin{eqnarray}\label{label2.1}
&&<out; \pi^+(q) D^0(p)|D^*(Q); in> =\\
&&=i(2\pi)^4\delta^{(4)}(Q-p-q)\frac{M(D^{*+}(Q)\to D^0(p)\pi^+(q))}{(2
E_{\pi^+}V 2E_{D^0}V 2E_{D^{*+}}V)^{1/2}}\nonumber
\end{eqnarray}
with
\begin{eqnarray}\label{label2.2}
M(D^{*+}(Q)\to D^0(p)\pi^+(q)) = g_{D^{\ast}D\pi}\,q\cdot e(Q),
\end{eqnarray}
where $E_i$ ($i=\pi^+, D^0$ or $D^{*+}$) is the energy of the $i$--meson,
and $V$ is the normalization volume, then $e^{\nu}(Q)$ is the 4--vector of
the $D^{*+}$--meson polarization. Note that by definition the coupling
constant $g_{D^{\ast}D\pi}$ in Eq.(\ref{label2.2}) involves a factor 2 with
respect to that defined in Ref.[1] (see Eq.(8) of Ref.[6]).

The computation of the matrix element $M(D^{*+}(Q)\to D^0(p)\pi^+(q))$ at
leading order in large $N$, number of quark colours, and $M_c$ expansion
has been carried out in Ref.[6]. The result reads
\begin{eqnarray}\label{label2.3}
M(D^{*+}(Q)\to D^0(p) \pi^+(q)) = \sqrt{2} g_{\pi^+} \frac{\sqrt{M_D
M_{D^*}}}{\bar{v}^{\prime}} {\ell n}\Bigg(\frac{\bar{v}^{\prime}}{4m}\Bigg)
q\cdot e(Q).
\end{eqnarray}
This gives one the coupling constant
\begin{eqnarray}\label{label2.4}
g_{D^{*+}D^0\pi^+}=g_{D^{*}D\pi}=\frac{4\pi\sqrt{2}}{\sqrt{N}}\frac{\sqrt{M_D M_
{D^*}}}{\bar{v}^{\prime}} {\ell n}\Bigg(\frac{\bar{v}^{\prime}}{4m}\Bigg) =
5.2,
\end{eqnarray}
where we have set $M_c=M_D=1.86\;{\rm GeV}$ and $M_{D^*}=2.00\;{\rm GeV}$
[1,2], and  $g_{\pi^+}=4\pi/\sqrt{N}$ [6,11--16] at $N=3$, then
$\bar{v}^{\prime}=4\Lambda=2.66\;{\rm GeV}$ [6,11--16] and $\Lambda$ is the
cut--off in Euclidean 3--momentum space connected with $\Lambda_{\chi}$,
the scale of spontaneous breaking of chiral symmetry (SB$\chi$S) in
(CHPT)$_q$, via the relation $\Lambda=\Lambda_{\chi}/\sqrt{2}=0.67\;{\rm
GeV}$ at $\Lambda_{\chi}=0.94\;{\rm GeV}$ [10]. The cut--off $\Lambda$
appears due to the computation of heavy--light constituent quark loops
describing strong low--energy interactions in HQET and (CHPT)$_q$  at
scales $p < \Lambda_{\chi}$.

The coupling constant $g_{\bar{B}^*\bar{B}\pi}$ can be obtained from
Eq.(\ref{label2.4}) by the change $M_c\to M_b$, $M_D\to M_{\bar{B}}$ and
$M_{D^*}\to M_{\bar{B}^*}$
\begin{eqnarray}\label{label2.5}
g_{\bar{B}^*\bar{B}\pi}=\frac{4\pi\sqrt{2}}{\sqrt{N}}\frac{\sqrt{M_{\bar{B}} M_{
\bar{B}^*}}}{\bar{v}^{\prime}} {\ell
n}\Bigg(\frac{\bar{v}^{\prime}}{4m}\Bigg)  =14.3,
\end{eqnarray}
where we have set $M_{\bar{B}^*}=M_{\bar{B}}=M_b=5.3\;{\rm GeV}$  [6].

Our predictions for the coupling constants $g_{D^{*}D\pi}=5.2$ and
$g_{\bar{B}^*\bar{B}\pi}=14.3$ agree with those given in Refs.[2,3].

\section{Coupling constants to next--to--leading order in heavy quark mass
expansion}
\setcounter{equation}{0}

\hspace{0.2in} Now we should proceed to the analysis of next--to--leading
order corrections in heavy quark mass expansion. We suggest to start with
the computation of the $1/M_c$ corrections to the coupling constant
$g_{D^{*}D\pi}$. In our approach, where the amplitudes are described in
terms of constituent quark loops with virtual momenta restricted by the
SB$\chi$S scale $\Lambda_{\chi}\ll M_c$, $1/M_c$ corrections can be
computed by direct expansion of the $c$--quark Green function in the
momentum representation [15,16]. Since the virtual momentum $k$ is
restricted by the SB$\chi$S scale $\Lambda_{\chi}\ll M_c$, we can expand
the $c$--quark Green function in powers of $k/M_c$ and, holding only the
terms of order $O(1/M_c)$, obtain [15,16]
\begin{eqnarray}\label{label3.1}
&&\frac{1}{M_c-\hat{k}-\hat{p}}=
-\Bigg(\frac{1+\hat{v}}{2}\Bigg)\,\frac{1}{v\cdot k + i\,0} - \nonumber\\
&&-\frac{1}{2M_c}\,\Bigg[\frac{\hat{k}}{v\cdot k + i\,0} -
\Bigg(\frac{1+\hat{v}}{2}\Bigg)\,\frac{k^2}{(v\cdot k + i\,0)^2}\Bigg] +
O\Bigg(\frac{1}{M^2_c}\Bigg),
\end{eqnarray}
where, as usually, we have set $p^{\mu}=M_D v^{\mu}=M_c v^{\mu}$. Thus, the
amplitude $M(D^{*+}(Q)\to D^0(p) \pi^+(q))$ can be given by
\begin{eqnarray}\label{label3.2}
&&M(D^{*+}(Q)\to D^0(p) \pi^+(q)) = \nonumber\\
&&=M^{(0)}(D^{*+}(Q)\to D^0(p)\pi^+(q)) + M^{(1)}(D^{*+}(Q)\to D^0(p) \\
\pi^+(q)),
\end{eqnarray}
where the term $M^{(0)}(D^{*+}(Q)\to D^0(p) \pi^+(q))$ is given by
Eq.(\ref{label2.3}), while $M^{(1)}(D^{*+}(Q)\to D^0(p) \pi^+(q))$ reads
\begin{eqnarray}\label{label3.3}
&&M^{(1)}(D^{*+}(Q)\to D^0(p) \pi^+(q)) =
g_D\,g_{D^*}\,\frac{g_{\pi^+}}{\sqrt{2}}\,\frac{1}{2
M_c}\frac{N}{16\pi^2}\,e^{\mu}(Q)\int\frac{d^4k}{\pi^2i}\nonumber\\
&&{\rm tr}\Bigg\{\frac{1}{m-\hat{k}+\hat{q}}\gamma^5
\frac{1}{m-\hat{k}}\gamma^5\Bigg[\frac{\hat{k}\gamma_{\mu}}{v\cdot k +
i\,0} - \Bigg(\frac{1+\hat{v}}{2}\Bigg)\gamma_{\mu}\frac{k^2}{(v\cdot k +
i\,0)^2}\Bigg]\Bigg\}.
\end{eqnarray}
After the calculation of the trace over Dirac matrices we reduce the r.h.s.
of Eq.(\ref{label3.3}) to the form
\begin{eqnarray}\label{label3.4}
&&M^{(1)}(D^{*+}(Q)\to D^0(p) \pi^+(q))=
g_D\,g_{D^*}\,\frac{g_{\pi^+}}{\sqrt{2}}\,\frac{1}{2
M_c}\,\frac{N}{16\pi^2}\,e^{\mu}(Q)\,\times\nonumber\\
&&\Bigg\{\int\frac{d^4k}{\pi^2i}\,\frac{4(m^2-k^2)\,k_{\mu}+2\,k^2\,q_{\mu}}{[m^
2-(k-q)^2 -i\,0][m^2-k^2 -i\,0]}\,\frac{1}{[v\cdot k + i\,0]} - \nonumber\\
&&- \int\frac{d^4k}{\pi^2i}\,\frac{2\,m\,k^2\,q_{\mu}}{[m^2-(k-q)^2
-i\,0][m^2-k^2 -i\,0]}\,\frac{1}{[v\cdot k + i\,0]^2}\Bigg\}.
\end{eqnarray}
Following the prescription of applied in our effective quark model
[6,10--16] (see also Ref. [17]) for the calculation of constituent quark
loop diagrams we have held only the leading divergent contributions
dropping out the contributions independent of the cut--off. This
prescription is nothing more than a trivial phenomenological description of
quark confinement. Since the appearance of free quarks in the intermedite
states of low--energy hadron interactions described by the constituent
quark loop diagrams with constant constituent quark masses is related to
the imaginary parts of these diagrams, the trivial prohibition of such an
appearance is to keeping the divergent contributions of quark loop diagrams
related to the real parts of quark diagrams. Moreover, the divergent parts
of quark diagrams, being polynomials in powers of the momenta of
interacting hadrons, provide local effective low--energy interactions of
hadrons in agreement with effective Chiral Lagrangian approach. In terms of
the cut--off and the constituent quark mass this procedure restores fully
all effective coupling constants of low--energy interactions of low--lying
mesons represented at the hadron level by effective Chiral Lagrangians
[10].

The integral containing $k_{\mu}$ in the integrand has the following
Lorentz structure: $A(v\cdot q,q^2)\,v_{\mu} + B(v\cdot q,q^2)\,q_{\mu}$.
The contribution of the term linear in $v_{\mu}$ is of order $O(1/M^2_c)$
due to $Q_{\mu} e^{\mu}(Q)=0$ and should be dropped out. Therefore, only
the terms linear in $q_{\mu}$ should be held and give the contributions
leading in chiral expansion. The integration over $k$ gives one
\begin{eqnarray}\label{label3.5}
&&\int\frac{d^4k}{\pi^2i}\frac{4(m^2-k^2)k_{\mu}+2k^2q_{\mu}}{[m^2-(k-q)^2
-i0][m^2-k^2 -i0]}\frac{1}{[v\cdot k + i0]}
=-\frac{2}{3}\,\bar{v}^{\prime}\,q_{\mu},\nonumber\\
&&\int\frac{d^4k}{\pi^2i}\frac{2mk^2q_{\mu}}{[m^2-(k-q)^2 -i0][m^2-k^2
-i0]}\frac{1}{[v\cdot k + i0]^2} = \nonumber\\
&&\hspace{1in} = -8\,m\,{\ell n}\Bigg(\frac{\bar{v}^{\prime}}{4
m}\Bigg)\,q_{\mu}.
\end{eqnarray}
For the computation of the intergals in Eq.(\ref{label3.5}) we have used
the integrals
\begin{eqnarray}\label{label3.6}
\int\frac{d^4k}{\pi^2i}\,\frac{k_{\mu}k_{\nu}}{[m^2-k^2
-i\,0]^2}\,\frac{1}{[v\cdot k + i\,0]} &=&
\frac{2}{3}\,\bar{v}^{\prime}\,(g_{\mu\nu} - v_{\mu}v_{\nu}),\nonumber\\
\int\frac{d^4k}{\pi^2i}\,\frac{1}{[m^2-k^2 -i\,0]}\,\frac{1}{[v\cdot k +
i\,0]} &=& -\,\bar{v}^{\prime},\nonumber\\
\int\frac{d^4k}{\pi^2i}\,\frac{1}{[m^2-k^2 -i\,0]}\,\frac{1}{[v\cdot k +
i\,0]^2} &=& 4\,{\ell n}\Bigg(\frac{\bar{v}^{\prime}}{4 m}\Bigg).
\end{eqnarray}
We should accentuate that no logarithmically divergent contributions have
been omitted in the linearly divergent momentum integrals. Indeed, for
example, one can show that
\begin{eqnarray}\label{label3.7}
&&\int\frac{d^4k}{\pi^2i}\,\frac{1}{[m^2-k^2 -i\,0]}\,\frac{1}{[v\cdot k +
i\,0]} = - \frac{1}{\pi}\,\int \frac{d^3k}{\vec{k}^{\,2} + m^2} =\nonumber\\
&&= - 4\,\Lambda + 4\,m\,{\rm arctg}\frac{\Lambda}{m}\to - \bar{v}^{\prime}
+ 2\,\pi\,m.
\end{eqnarray}
Thus, following our phenomenological description of confinement, demanding
to drop out the terms independent of the cut--off, and collecting all
divergent terms of order $O(1/M_c)$ we obtain
\begin{eqnarray}\label{label3.8}
&&M^{(1)}(D^{*+}(Q)\to D^0(p) \pi^+(q)) = \nonumber\\
&&=\sqrt{2}\,g_{\pi^+}\,\frac{\sqrt{M_D M_{D^*}}}{\bar{v}^{\prime}}
\Bigg[\frac{m}{M_c}{\ell n}\Bigg(\frac{\bar{v}^{\prime}}{4 m}\Bigg) -
\frac{\bar{v}^{\prime}}{12 M_c}\Bigg]\,q\cdot e(Q).
\end{eqnarray}
This gives the coupling constant $g_{D^* D\pi}$ to next--to--leading order
in large $M_c$ expansion
\begin{eqnarray}\label{label3.9}
g_{D^* D\pi} &=& \sqrt{2}\,g_{\pi^+}\,\frac{\sqrt{M_D
M_{D^*}}}{\bar{v}^{\prime}}\Bigg\{{\ell n}\Bigg(\frac{\bar{v}^{\prime}}{4
m}\Bigg) + \Bigg[\frac{m}{M_c}{\ell n}\Bigg(\frac{\bar{v}^{\prime}}{4
m}\Bigg) - \frac{\bar{v}^{\prime}}{12 M_c}\Bigg]\Bigg\}\nonumber\\
&=& 5.2\times (1 + 7.3\times 10^{-3}) = 5.24.
\end{eqnarray}
Thus, we have found that $1/M_c$ corrections to the constant $g_{D^* D\pi}$
make up less than 1$\%$ and can be neglected. The accounting of the
logarithmically divergent constributions together with the linearly ones
does not go beyond the scape of the approach. Indeed, for the correct
calculation of the kinetic terms of the low--lying scalar mesons related to
the kinetic terms of the low--lying pseudoscalar ones via chiral
transformations one needs to take into account both quadratically and
logarithmically divergent constributions [10].

For the computation of the coupling constant $g_{\bar{B}^*\bar{B}\pi}$ we,
following the procedure expounded above, get
\begin{eqnarray}\label{label3.10}
g_{\bar{B}^*\bar{B}\pi} &=& \sqrt{2} \,g_{\pi^+} \,
\frac{\sqrt{\displaystyle M_{\bar{B}}
M_{\bar{B}^*}}}{\bar{v}^{\prime}}\Bigg\{{\ell
n}\Bigg(\frac{\bar{v}^{\prime}}{4 m}\Bigg) + \Bigg[\frac{m}{M_b}{\ell
n}\Bigg(\frac{\bar{v}^{\prime}}{4 m}\Bigg) - \frac{\bar{v}^{\prime}}{12
M_b}\Bigg]\Bigg\}\nonumber\\
&\simeq& 14.3,
\end{eqnarray}
where the numerical value is obtained at  $M_{\bar{B}^*} = M_{\bar{B}} =
M_b = 5.3\,{\rm GeV}$ [1]. Thus, we predict negligibly small
next--to--leading corrections in heavy $M_b$ expansion to the coupling
constant $g_{\bar{B}^*\bar{B}\pi} = 14.3$.

\section{Conclusion}

When applying the effective quark model with chiral $U(3)\times U(3)$
symmetry incorporating HQEF and (CHPT)$_q$ to the computation the coupling
constants $g_{D^* D\pi}$ and $g_{\bar{B}^*\bar{B}\pi}$ to next--to--leading
order in heavy--quark mass expansion, we have computed at leading order in
large N expansion and in the chiral limit the numerical values
$g_{D^{*}D\pi}=5.2$ and $g_{\bar{B}^*\bar{B}\pi} =14.3$ compared reasonably
well with the results obtained by  Colangelo et al. [2] within QCD sum
rules approach and by Dosch and Narison [3] within {\it full} QCD. Then, we
have shown that next--to--leading corrections in large heavy--quark mass
expansion give negligible contributions and the main values of the coupling
constants are induced at leading order. Our results as well as those
predicted by Colangelo et al. [2] and Dosch and Narison [3] are a factor of
2 less compared with those obtained by Belyaev et al.[4]. Therefore, in
comparison with Ref.[4] we predict much less contribution of vector mesons
$D^*$ and $\bar{B}^*$ to the form factors of the semileptonic decays $D\to
\pi\ell\nu_{\ell}$ and $\bar{B}\to \pi\ell\nu_{\ell}$. The former is very
important for the understanding of a mechanism of the semileptonic decays
$D\to \pi\ell\nu_{\ell}$ and $\bar{B}\to \pi\ell\nu_{\ell}$. It should be
stressed that our numerical results contain a theoretical uncertainty $\pm
\Delta$, the rough estimate of which gives $\Delta \simeq 50\%$ [6].
However for all apllications of HQET and (CHPT)$_q$ [6,11--16] the
resultant agreements between theoretical and experimental data turn out
much better.

\section{Acknowledgement}

The authors would like to thank President Professor Abdus Salam, the
International Atomic Energy Agency and UNESCO for hospitality at the
International Centre for Theoretical Physics, Trieste, where the main part
of this paper has been carried out.

\end{document}